# Development and experimental validation of an in-house treatment planning system with greedy energy layer optimization for fast IMPT

Aoxiang Wang[1,2], Ya-Nan Zhu[2*], Jufri Setianegara[2], Yuting Lin[2], Peng Xiao[1], Qingguo Xie[1*], and Hao Gao[2]

[1]Department of Biomedical Engineering, Huazhong University of Science and Technology, Wuhan, China

[2]Department of Radiation Oncology, University of Kansas Medical Center, USA

**Email:** yzhu4@kumc.edu , qgxie@hust.edu.cn

**Acknowledgment**

The authors are very thankful to the valuable comments from reviewers. This research is partially supported by the NIH grants No. R37CA250921, R01CA261964, and a KUCC physicist-scientist recruiting grant.

**Abstract.**

**Background:** Intensity-modulated proton therapy (IMPT) using pencil beam technique scans tumor in a layer by layer, then spot by spot manner. It can provide highly conformal dose to tumor targets and spare nearby organs-at-risk (OAR). Fast delivery of IMPT can improve patient comfort and reduce motion-induced uncertainties. Since energy layer switching time dominants the plan delivery time, reducing the number of energy layers is important for improving delivery efficiency. Although various energy layer optimization (ELO) methods exist, they are rarely experimentally validated or clinically implemented, since it is technically challenging to integrate these methods into commercially available treatment planning system (TPS) that is not open-source.

**Purpose:** This work develops and experimentally validates an in-house TPS (IH-TPS) that incorporates a novel ELO method for the purpose of fast IMPT.

**Methods:** The dose calculation accuracy of IH-TPS is verified against the measured beam data and the RayStation TPS. For treatment planning, a novel ELO method via greed selection algorithm is proposed to reduce energy layer switching time and total plan delivery time. To validate the planning accuracy of IH-TPS, the 3D gamma index is calculated between IH-TPS plans and RayStation plans for various scenarios. Patient-specific quality-assurance (QA) verifications are conducted to experimentally verify the delivered dose from the IH-TPS plans for several clinical cases.

**Results:** Dose distributions in IH-TPS matched with those from RayStation TPS, with 3D gamma index results exceeding 95% (2mm, 2%). The ELO method significantly reduced the delivery time while maintaining plan quality. For instance, in a brain case, the number of energy layers was reduced from 78 to 40, leading to a 62% reduction in total delivery time. Patient-specific QA validation with the IBA Proteus®ONE proton machine confirmed a >95% pass rate for all cases.

**Conclusions:** An IH-TPS equipped with a novel ELO algorithm is developed and experimentally validated for the purpose of fast IMPT, with enhanced delivery efficiency and preserved plan quality.

# 1. Introduction

Intensity-modulated proton therapy (IMPT) based on the pencil beam scanning (PBS) technique can deliver highly conformal dose to tumor targets, effectively reducing the exposure of surrounding healthy tissues when compared to photon therapy or passive scattering proton therapy [1]. The fast treatment delivery of IMPT has several benefits, such as enhanced patient comfort, reduced motion impact [2,3], and increased patient throughput, allowing more efficient use of limited proton treatment resources and expanding access of proton therapy [4,5].

The IMPT treatment scans the tumor target layer by layer, and then spot by spot [6,7]. The delivery time of PBS based IMPT mainly consists of three parts: dose delivery time, spot scanning time, and energy layer switching time (ELST). The ELST constitutes a significant portion of the overall plan delivery. For instance, in cyclotron-based and synchrotron-based systems like the IBA's proton machines, the average ELST accounts for more than half of the total plan delivery time [8,9]. This motivates the reduction of energy switching time in the pursuit of delivery efficiency.

Several energy layer optimization (ELO) methods have been proposed to reduce the energy switching time. [10] minimized the logarithm of total spot weight per energy layer and then excluded the low-weight energies. [11] formulated the problem as a mixed integer problem and iteratively deleted one energy at a time. [12] introduced the root mean squared regularizer to reduce irrelevant energies. [13] proposed MMSEL, which addressed energy reduction through group sparsity and minimum monitor unit constraints. [14] developed so-called the CARD method, applying the cardinality-sparsity constraint to the energy vector. However, these ELO methods are rarely experimentally validated or clinically implemented, as it is technically challenging to integrate these methods into commercially available treatment planning system (TPS) that is not open-source.

This work aims to develop and experimentally validate an in-house TPS (IH-TPS) with novel ELO capability for the purpose of fast IMPT. The equipped novel ELO method directly hard-constrain the number of proton energies and selects the proper energies using a greedy method originated from compressive sensing. Experimental validation will be performed to demonstrate the deliverability and the accuracy of the IH-TPS plans, in comparison to the RayStation plans and delivered plans on a clinical proton machine.

## 2. Methods

*2.1. IH-TPS: dose engine*

The dose engine of IH-TPS is based on a modified version of pencil beam dose calculation algorithm [15]. To model our IBA's Proteus®ONE proton machine, two source-to-axis distances are used. To improve the accuracy of dose calculation in patient, we convert the Hounsfield Unit values to electron densities using CT calibration measurements obtained from our CT scanner.

In order to ensure the clinical deliverability and accuracy of IH-TPS, the machine parameters and the configuration of IH-TPS are matched to our proton machine. The beam model consists of the parameters from all energy layers. To create the beam data for IH-TPS, RayStation's Monte Carlo (MC) dose engine is used to simulate the 3D dose of different energy layers in a 40×40×40 cm³ water phantom. The base data consists of single-spot dose results at 2.5 MeV intervals, ranging from 70 to 225 MeV. Each energy layer is delivered as a single spot with one monitor unit, with the beam isocenter set at the surface of the water phantom. These base data are used to calculate the beam parameters for all energy layers, including integrated depth dose (IDD), range, peak, and spot sizes in water at different depths. By adjusting these beam parameters, a new beam model of 63 energy layers is developed for IH-TPS.

*2.2 IH-TPS: inverse planning method*

*2.2.1 Energy layer optimization*

The inverse planning of IH-TPS with ELO is to solve the following optimization problem

$$\min_{x \in \mathfrak{R}^n} f(d, \Omega)$$
$$\text{s.t.} \begin{cases} d = Dx \\ x_j \in \{0\} \cup [G_{\min}, +\infty), j \leq n \end{cases} \quad (1)$$

where $D \in \mathbb{R}^{m \times n}$ is dose influence matrix and $x \in \mathbb{R}^n$ is beam intensity. The variable $x$ satisfies the minimum-monitor unit constraint (MMU) with MMU threshold $G_{min}$. $f(d,\Omega)$ is dose-volume histogram (DVH) [16,17] related plan objective, enforcing the consistency between the prescribed dose and optimized dose $d$. Based on quadratic model, the $f(d,\Omega)$ has the following form

$$f(d,\Omega) = \sum_{i=1}^{N_{L2}} \omega_{1,i} \left\| d_{\Omega_{1,i}} - d_{1,i} \right\|^2 + \sum_{i=1}^{N_{DVH-max}} \omega_{2,i} \left\| d_{\Omega_{2,i}} - d_{2,i} \right\|^2 + \sum_{i=1}^{N_{DVH-min}} \omega_{3,i} \left\| d_{\Omega_{3,i}} - d_{3,i} \right\|^2 \qquad (2)$$

The first term is the $L_2$ type objective that sums the squared deviation between optimized dose and prescription dose over voxels under consideration. The second term is DVH-max constraint objective that limits the upper bound of ratio of overdosed voxels in region-of-interest (ROI). The third term is DVH-min constraint objective that sets the lower bound of ratio of underdose voxels inside the target. $\omega_{1,i}$, $\omega_{2,i}$ and $\omega_{3,i}$ are positive objective weights and $\Omega$ is the indexes of voxels that violate the DVH constraints [16,17].

The number of columns in $D$ is determined by the number of spots for each energy layer, the number of energy layers for each delivery angle, and the total number of delivery angles. As previously mentioned, the number of energy layers plays a crucial role in minimizing the overall delivery time. Current treatment planning systems typically utilize all available proton energies to cover the target. However, redundancies between energy layers may exist, meaning that an acceptable plan quality can be achieved with fewer energy layers. Thus, to improve delivery efficiency, it is essential to reduce the number of energy layers.

Firstly, the influence matrix $D$ can be decomposed by the energy variable as the follows

$$D = [D_1, D_2, \cdots, D_{NE}] \qquad (3)$$

where $D_i$ is the influence matrix for $i^{th}$ energy, and the number of columns of $D_i$ is determined by the number of spots. The beam intensity $x$ can be similarly decomposed by

$$x = [x_1, x_2, \cdots, x_{NE}]^T \qquad (4)$$

Each $x_i$ represents a given block of $x$, and their indexes are given by

$$[1,2,\cdots,N] = \Lambda = [\Lambda_1, \Lambda_2, \cdots, \Lambda_{NE}] \tag{5}$$

Limiting the number of energies is equivalent to restricting the number of nonzero blocks of *x* in (4). Denote the $\|x\|_{BS}$ as the number of nonzero blocks, then the energy layer constraint can be modeled as

$$\|x\|_{BS} \leq N \tag{6}$$

where $N < N_E$ is predefined number of energies to be used.

Combing (1) and (6), we get the following energy layer optimization model

$$\min_{x \in \eth^n} f(d, \Omega)$$
$$s.t. \begin{cases} d = Dx \\ x_j \in \{0\} \cup [G_{\min}, +\infty), j \leq n \\ \|x\|_{BS} \leq N \end{cases} \tag{7}$$

### 2.2.2 Optimization algorithm

Eq. (7) can be solved by classical first-order methods, e.g., the primal-dual method [18], alternating direction method of multipliers (ADMM) [19-24], etc. As the problem is highly nonconvex, these methods can be easily trapped in local minimal. Note that the main challenge in solving (7) lies in selecting the appropriate N blocks. Inspired by methods in compressive sensing [25], we can address this challenge by selecting these blocks in a greedy fashion. Without loss of generality, let us temporarily set aside the DVH plan objectives, MMU constraint, and only consider the L2 type objective in (2). Rewriting the L2 type objective in (2) concisely into normed squared, (7) can be reformulated as

$$\min_{\|x\|_{GS} \leq N} \|Ax - b\|_2^2 \tag{8}$$

Analogous to *D,* the *A* can be decomposed *as*

$$A = [A_1, A_2, \cdots, A_{NE}] \tag{9}$$

Finding proper energies is transferred to find proper nonzero blocks of *x* such that (8) is minimized. The (8) is exactly same as the classical compressive sensing problem with *N* block sparsity constraint. Here we will develop orthogonal matching pursuit method [26,27] to solve this optimization problem, which goes as the follows

**Input:** $A, b$ and $N$.

**Initialize:** $x^0 = \mathbf{0}, S^0 = \emptyset$.

**For** $k = 0, 1, ..., N-1$

$$j^{k+1} = \arg\max_j \left\{ \left\| A_j^T (b - Ax^k) \right\|_2 \right\} \quad (10a)$$

$$S^{k+1} = S^k \cup \{\Lambda_{j^{k+1}}\}; \quad (10b)$$

$$x^{k+1} = \arg\min_{\text{supp}(z) \in S^{k+1}} \left\| b - Az \right\|_2^2 \quad (10c)$$

**End**

**Input:** $x^N$

In each step, the ELO method selects the block that is mostly related to current residual and adds to the support set $S^k$ (10a-10b), then updates $x$ by projecting onto the span of the selected blocks (10c). This approach offers a greedy method for block selection. This block selection method can be utilized in solving (7) as follows

**Input:** $A, b, N, G_{\min}, Itr \leq N_E$.

**Initialize:** $x_1^0 = x_2^0 = \mathbf{0}, S_1^0 = S_2^0 = \emptyset$. The number of activated energy layers $C_E = 0$.

**For** $k = 0, 1, ..., Itr-1$

$C_E = C_E + 1;$

$$j^{k+1} = \arg\max_j \left\{ \left\| A_j^T (b - Ax_1^k) \right\|_2 \right\} \quad (11a)$$

$$S_1^{k+1} = S_1^k \cup \{\Lambda_{j^{k+1}}\}; \quad (11b)$$

$$x_1^{k+1} = \arg\min_{\text{supp}(z) \in S_1^{k+1}} \left\| b - Az \right\|_2^2 \quad (11c)$$

$$S_2^{k+1} = S_2^k \cup \{\Lambda_{j^{k+1}}\}; \qquad (11d)$$

$$(f^{k+1}, x_2^{k+1}) = \begin{cases} \arg\min_{x \in \mathbb{R}^n} f(Dx, \Omega) \\ \text{s.t. } \operatorname{supp}(x) \subset S_2^{k+1}, x_j \in \{0\} \cup [G_{\min}, +\infty), j \in \operatorname{supp}(x) \end{cases} \qquad (11e)$$

**If** $f^{k+1} \geq f^k$ $\qquad (11f)$

$$S_2^{k+1} = S_2^k \qquad (11g)$$

$$x_2^{k+1} = x_2^k \qquad (11h)$$

$$C_E = C_E - 1$$

**End**

**If** $C_E == N$

$$x^{out} = x_2^{k+1}$$

Break

**End**

**End**

**Output:** $x^{out}$

In each step, the generic greedy selection method (11a-11c) (on L2 objective) is employed to select candidate block (with indexes of the block added to the support set $S_1^k$). The selected energy maximally decreases the residual for (8) but may not be a good candidate for (7) due to the DVH-min, DVH-max plan objectives and MMU constraints. Thus, we solve (7) based on the selected blocks (with the updated support set $S_2^{k+1}$,) using iteration convex relaxation method [28-32] and ADMM and obtain its objective value $f^{k+1}$ (11e). The new block (as well as the updated $x_2^{k+1}$) is activated only if the $f^{k+1}$ is less than the optimal value of previous iteration (11f-11h). The iteration stops when the activated energy layers $C_E$ (the number of selected blocks in $S_2^k$) reaches the number $N$. During this iterative process, we can ensure that the objective function consistently decreases with each iteration, leading to an overall improvement in plan quality.

*2.3. Materials*

Three clinical cases, including abdomen, brain, and lung, were utilized to validate and evaluate the proposed IH-TPS in terms of both accuracy and efficiency. The dose influence matrix was generated on a 1 mm³ dose grid. Clinically used beam angles were used for each case: (90º, 180º, 270º) for the abdomen, (45º, 135º, 225º, 315º) for the brain, and (0º, 60º, 90º, 330º) for the lung.

Clinically-used DVH plan objectives were utilized, and all plans were normalized to D95% = 100% in CTV. The conformity index (CI) was evaluated, defined as $V_{100,\text{CTV}}^2/(V_{\text{CTV}} \times V_{100})$. ($V_{100,\text{CTV}}$: CTV volume receiving at least 100% of prescription dose; $V_{\text{CTV}}$: CTV volume; $V_{100}$: total body volume receiving at least 100% of prescription dose; ideally CI = 1). The dose quantities are in percentage with respect to the prescription dose; the values of objective value and CI are unitless.

The experimental validations were conducted: (1) proton spot weights and locations were scripted as a csv file to be imported to RayStation TPS for delivery on our proton machine and recalculating 3D dose in RayStation for comparison; (2) experimental validations were performed with patient-specific quality-assurance plan delivery in phantoms; (3) the absolute proton dose was measured via MatriXX ONE 2D ionization chamber array; (4) the gamma indexes was calculated to quantify the delivery accuracy, using the clinical standard for the passing criterion (≥95%, with 2%/2mm and a 10% threshold).

3. **Results**

*3.1. IH-TPS v.s. RayStation*

To compare IH-TPS and RayStation (which is clinically used at our institution), we conducted analyses under two scenarios: (1) to examine the IDD, central depth dose profile, lateral beam profile at beginning and peak, and three-dimensional gamma-analysis of a single spot irradiation in water; (2) to compare optimized plans (without ELO) for three clinical cases. For each scenario, three-dimensional gamma analysis [33], was employed with a distance to agreement (DTA) of 2 mm or 3 mm and a dose difference (DD) criterion of 2% or 3%. Only dose points exceeding 10% of the maximum dose value were included in the gamma analysis.

*3.1.1 Single spot*

IDD of single spot for three different energy layers (with peak positions at 5 cm, 15 cm, and 25 cm) were given in Figure1(a). The central depth dose profiles were given in Figure1(b). The lateral dose profiles were displayed (c) and (d). All the results were scaled to spread them for visualization and utilized identical image dimensions and dose resolutions (x = 1 mm, y = 1 mm, z = 1 mm). The results demonstrated a good agreement between IH-TPS and RayStation. The global gamma analyses (dta = 2 mm, Dd = 2%) pass rates for the three energy layers are 100%, 99.98%, and 96.51%, respectively.

*3.1.2 Different cases*

There were small differences between IH-TPS and RayStation as shown in Table 1. For instance, in terms of target coverage, the CI and target maximum dose for IH-TPS vs. RayStation were 0.90 and 110.8% vs. 0.84 and 112.9% for the abdomen, 0.72 and 113.1% vs. 0.64 and 116.1% for the brain, and 0.74 and 113.1% vs. 0.72 and 113.9% for the lung. Regarding OAR sparing, IH-TPS produced similar results to RayStation. For example, the mean kidney dose was 49.1% vs. 48.1% for the abdomen, the mean brainstem dose was 25.5% vs. 26.0% for the brain, and the mean esophagus dose was 16.9% vs. 15.4% for the lung. Additionally, the 3D gamma index results for the three cases were as follows: 95.59% (2mm, 2%) and 99.43% (3mm, 3%) for the abdomen; 97.09% (2mm, 2%) and 99.38% (3mm, 3%) for the brain; and 95.77% (2mm, 2%) and 99.45% (3mm, 3%) for the lung. The dose and DVH plots in Figures 2-4 further validate these results. Lastly, Figure 5 presents the QA results from one beam in each case. The gamma index was 97.2% (with 100% for the other two beams) for the abdomen, 100% (98.15%, 98.18%, 95.92% for the remaining beams) for the brain, and 100% (100%, 97.96%, and 97.56% for the other beams) for the lung.

*3.2. Energy layer optimization via IH-TPS*

We compared the results without (CONV) and with ELO (using the modified BOMP) in IH-TPS. The results with ELO required fewer energy layers to achieve similar plan quality, as shown in Table 2. For instance, with the target maximum dose kept at an acceptable level, ELO reduced the number of energy layers by 50% (from 50 to 25) for the abdomen, 62% (from 78 to 30) for the brain, and 30% (from 100 to 70) for the lung. In terms of target coverage, ELO showed comparable CI, such as 0.90 vs. 0.84 for the abdomen, 0.72 vs. 0.66 for the brain, and 0.74 vs. 0.70 for the lung. Additionally, ELO achieved similar OAR sparing, with mean kidney doses of 49.1% (CONV) vs. 50.3% (ELO) for the abdomen, mean brainstem dose of 25.5% vs. 29.7% for the brain, and mean esophagus dose of 16.9% vs. 18.2% for the lung. These findings are further validated by the dose plots in Figure 6 and DVH plots in Figure 7.

*3.3 Experimental validation of IH-TPS*

To demonstrate the effectiveness of the method, all newly generated plans using IH-TPS were recalculated in RayStation. As shown in Tables 3 and 4, the ELO method maintained its efficacy in RayStation. For instance, with the maximum target dose within 120% of prescription dose, the number of energy layers was reduced from 50 to 30 for the abdomen, leading to a decrease in delivery time from 110 seconds to 90 seconds. Similarly, for the brain, the number of energy layers was reduced from 78 to 40, shortening the delivery time from 105 seconds to 54 seconds. For the lung, energy layers decreased from 100 to 70, reducing the delivery time from 178 seconds to 127 seconds. Additionally, to verify the deliverability of the plans, we conducted QA and all cases exceeded the 95% pass rate. For the abdomen, beam 1 achieved 99.28%, beam 2 reached 100%, and beam 3 achieved 99.34%. For the brain, beam 1 and beam 2 both achieved 100%, while beam 3 and beam 4 attained 97.96% and 97.56%, respectively. For the lung, beam 1 achieved 98.99%, beam 2 and beam 3 both reached 100%, and beam 4 achieved 99.24%.

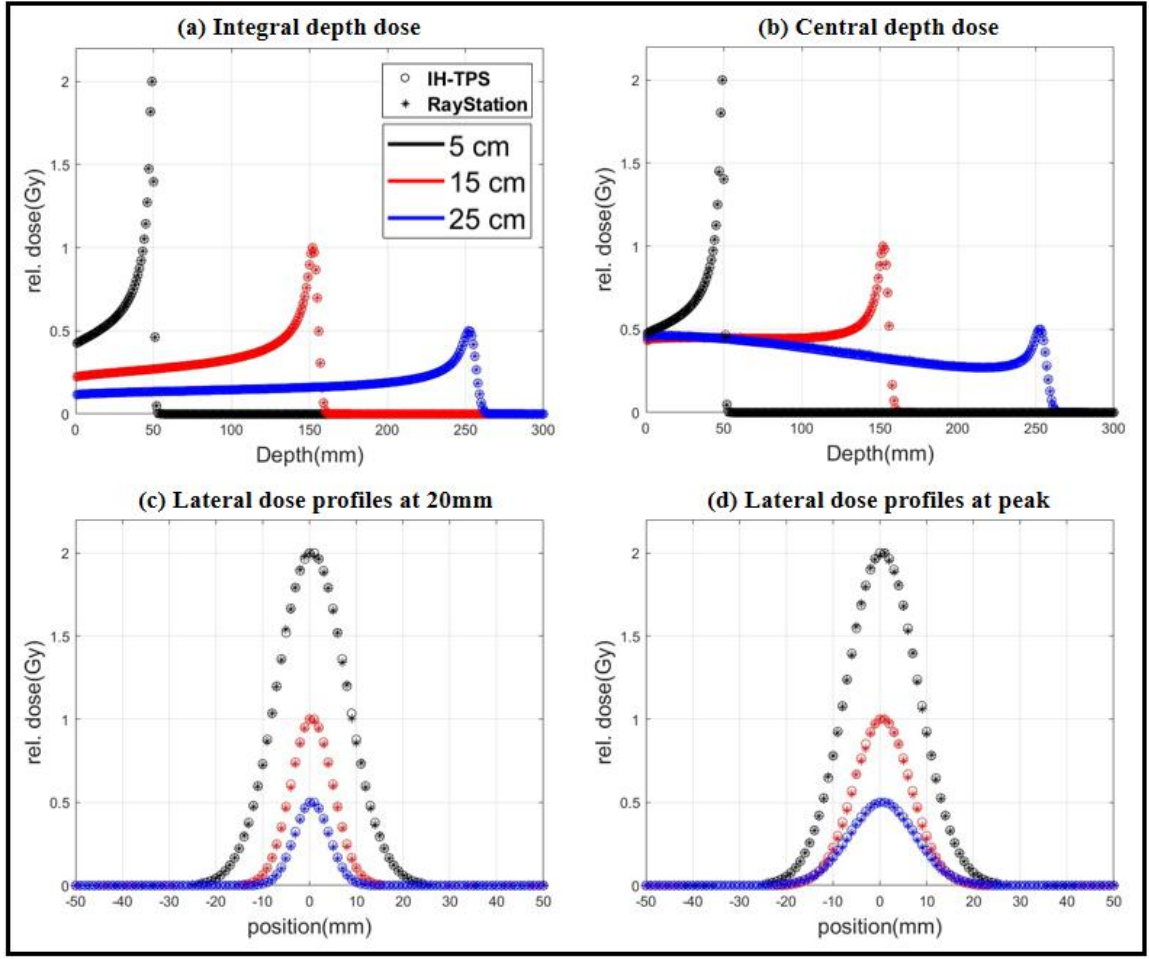

Figure 1. **Validation of beam model for IH-TPS.** Dose distributions of single spot for different energy layers with peak positions at 5 cm, 15 cm, and 25 cm. (a) Integral depth dose. (b) Central depth dose profiles. (c)-(d) Lateral dose profiles at the beginning (2 cm) and the peak. All the results were rescaled with maximum dose 2, 1, and 0.5.

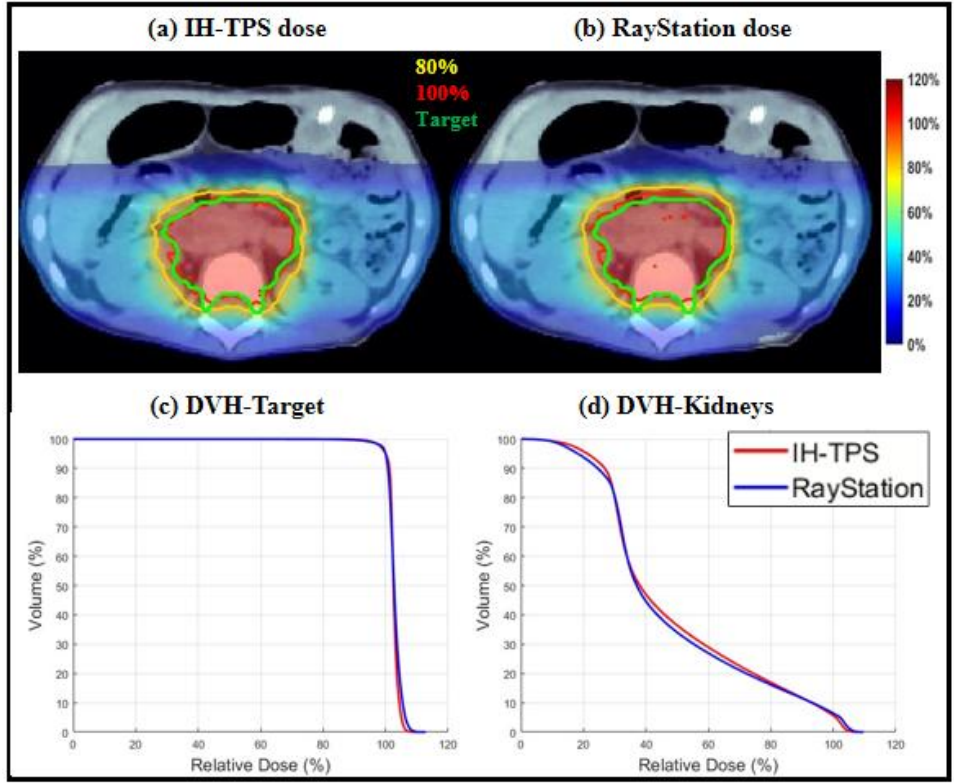

Figure 2. **Abdomen.** (a)-(b) Dose plots of IH-TPS and RayStation; (c)-(d) DVH plots of target and kidney. The dose plot window is [0%, 120%] of the prescription dose, with 80% (yellow) and 100% (red) isodose lines and CTV (green) highlighted in dose plots.

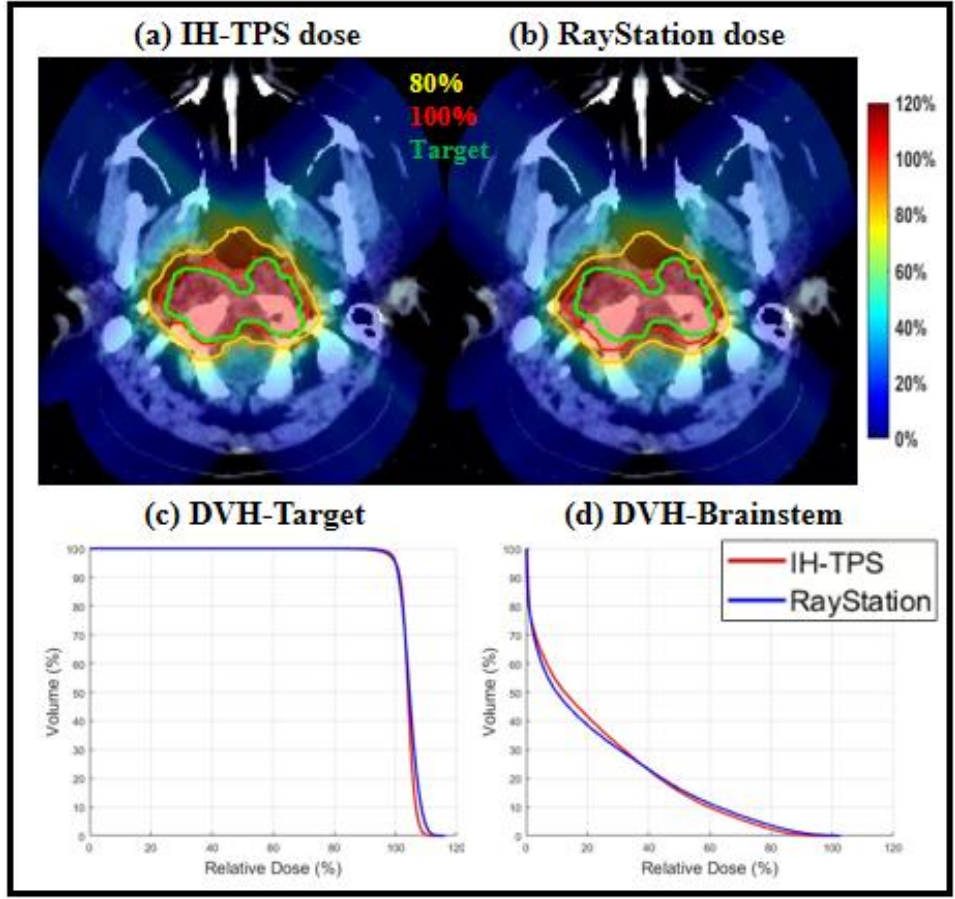

Figure 3. **Brain.** (a)-(b) Dose plots of IH-TPS and RayStation; (c)-(d) DVH plots of target and brainstem. The dose plot window is [0%, 120%] of the prescription dose, with 80% (yellow) and 100% (red) isodose lines and CTV (green) highlighted in dose plots.

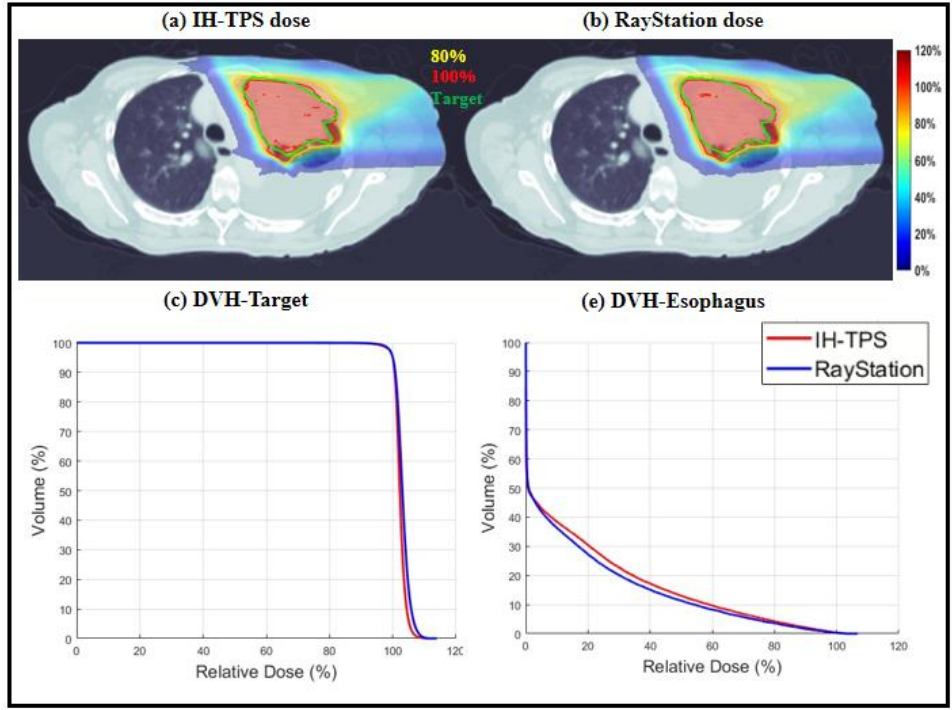

Figure 4. **Lung.** (a)-(b) Dose plots of IH-TPS and RayStation; (c)-(e) DVH plots of target and esophagus. The dose plot window is [0%, 120%] of the prescription dose, with 80% (yellow) and 100% (red) isodose lines and CTV (green) highlighted in dose plots.

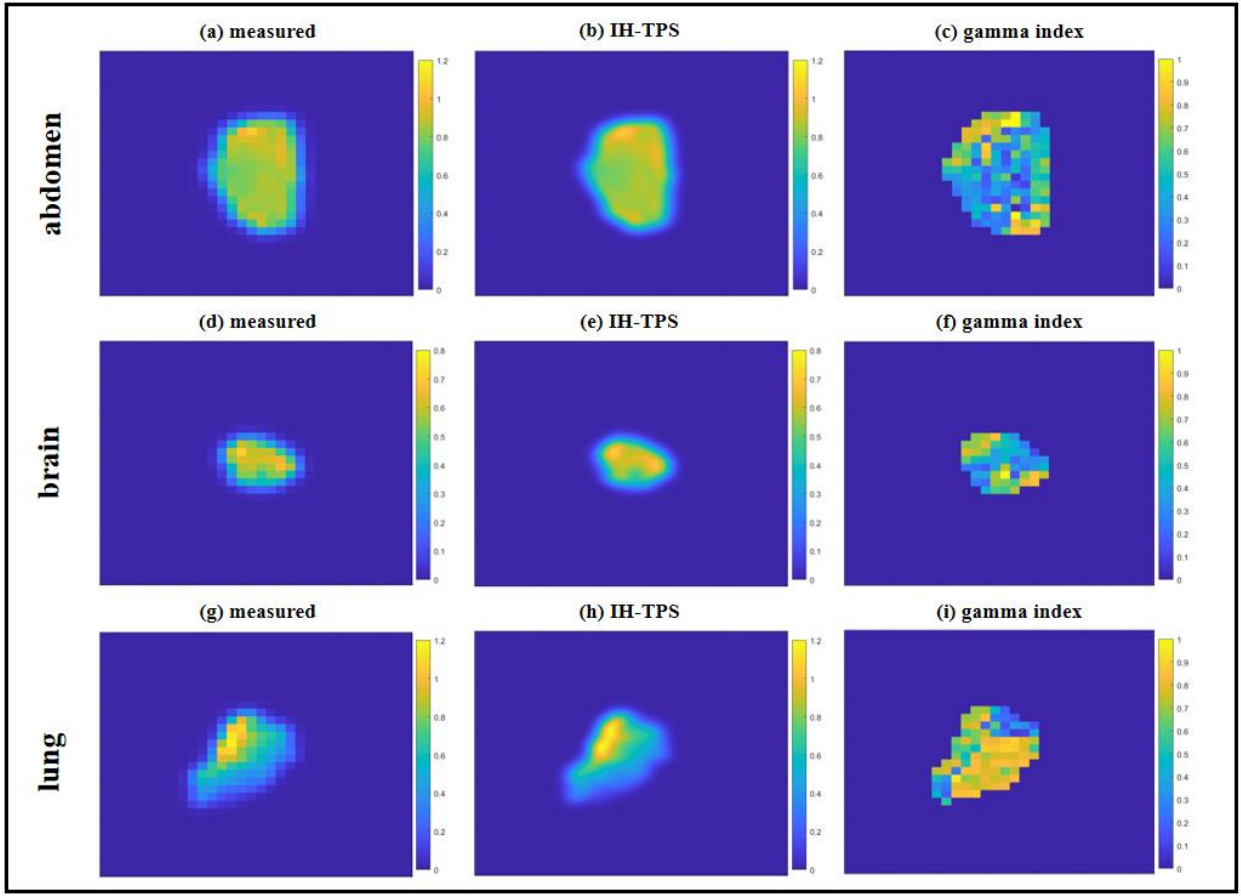

Figure 5. **QA results.** The measured ((a), (d), (g)), IH-TPS ((b), (e), (h)), and gamma index ((c), (f), (i)) for the abdomen, brain and lung, respectively.

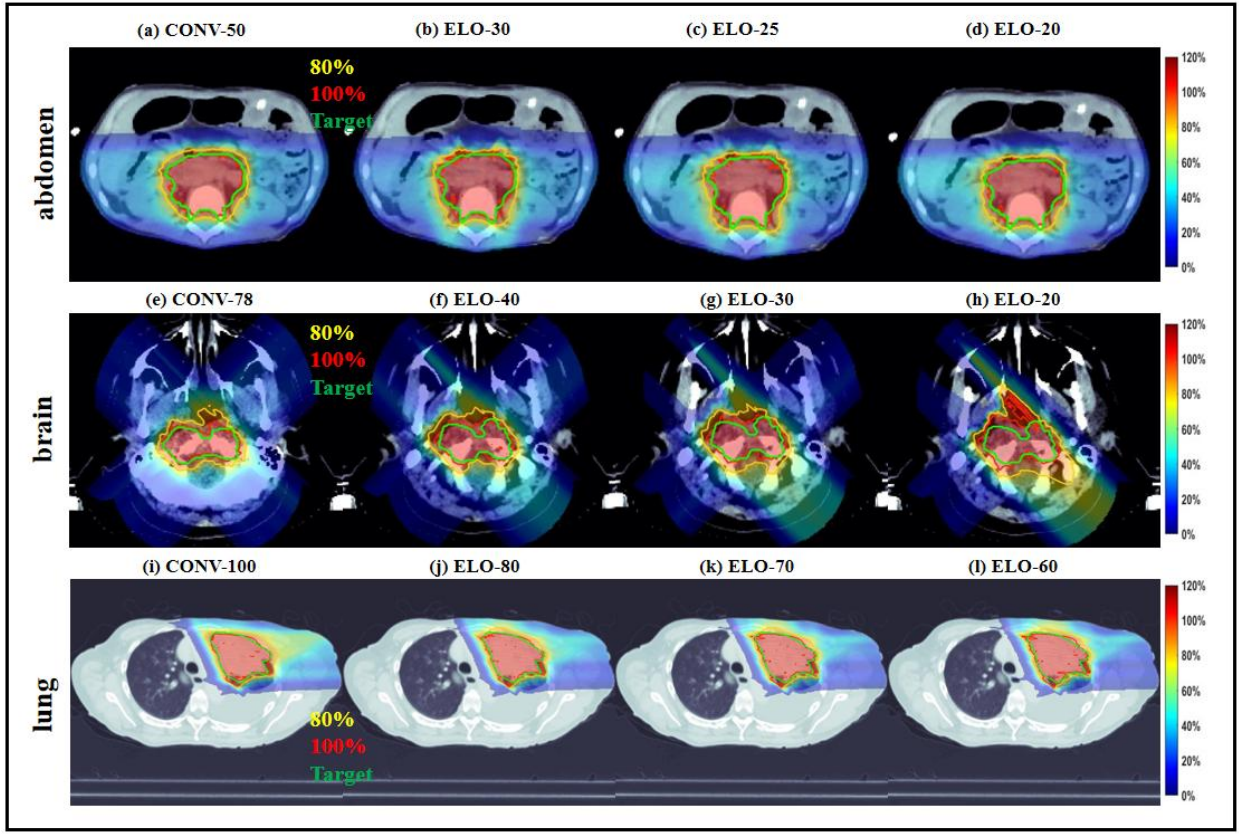

Figure 6. **Dose plots of CONV and ELO with different energy layers**. (a)-(d) Dose plots of CONV and ELO with different energy layers for the abdomen; (e)-(h) Dose plots of CONV and ELO with different energy layers for the brain; (i)-(l) Dose plots of CONV and ELO with different energy layers for the lung; The dose plot window is [0%, 120%] of the prescription dose, with 80% (yellow) and 100% (red) isodose lines and CTV (green) highlighted in dose plots.

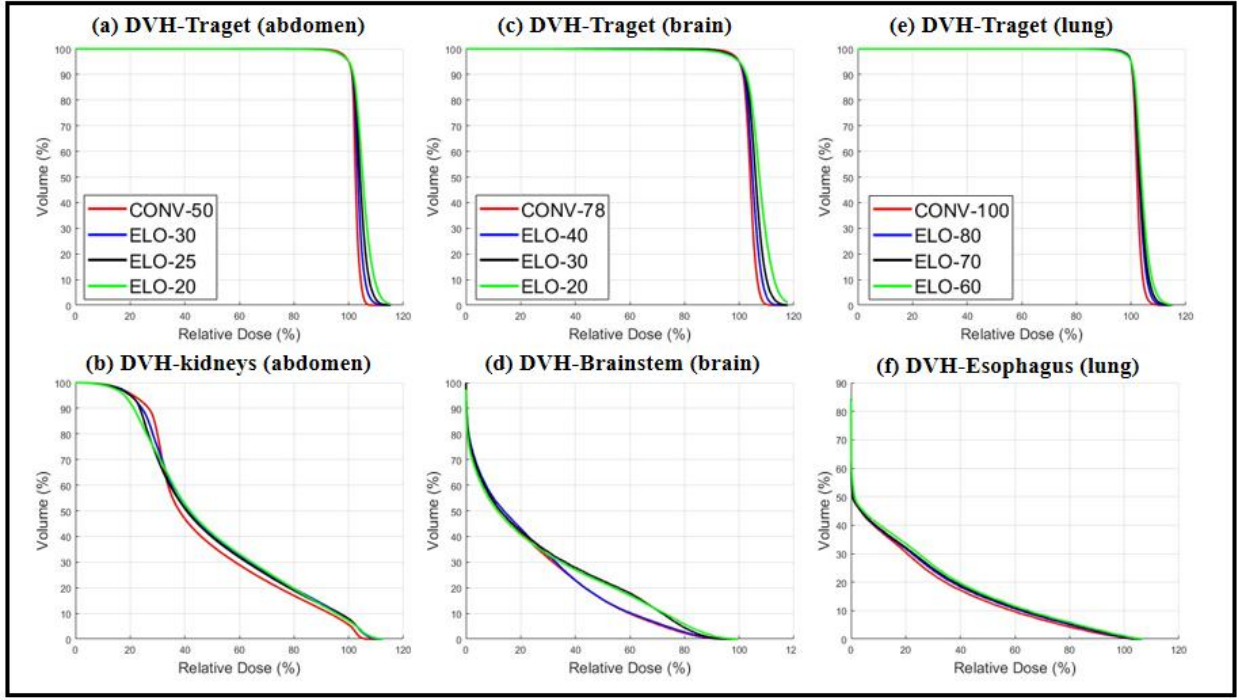

Figure 7. **DVH plots for CONV and ELO with different energy layers.** (a)-(b) The DVH plots of the target and kidneys for the abdomen. (c)-(d) The DVH plots of the target and brainstem for the brain. (e)-(f) The DVH plots of the target and esophagus for the lung.

Table 1. **IH-TPS** *v.s.* **RayStation.** The plan-quality dosimetric quantities from left to right: optimization objective value f, conformity index (CI), max target dose Dmax, mean OAR dose Dmean_oar and max OAR dose Dmax_oar.

| Case | Model | $f$ | CI | Dmax | Dmax_oar | Dmean_oar |
|---|---|---|---|---|---|---|
| **Abdomen** | IH-TPS | 2.23 | 0.90 | 110.8% | 108.1% | 49.1% |
| | RayStation | 2.42 | 0.84 | 112.9% | 109.6% | 48.1% |
| **Brain** | IH-TPS | 2.21 | 0.72 | 113.1% | 97.7% | 25.5% |
| | RayStation | 2.88 | 0.64 | 116.1% | 102.7% | 26.0% |
| **Lung** | IH-TPS | 1.15 | 0.74 | 113.1% | 106.8% | 16.9% |
| | RayStation | 1.32 | 0.72 | 113.9% | 104.0% | 15.4% |

Table 2. **CONV *v.s.* ELO plans optimized in IH-TPS.** The plan-quality dosimetric quantities from left to right: number of used energies, optimization objective value *f*, conformity index (CI), max target dose Dmax, mean OAR dose Dmean_oar and max OAR dose Dmax_oar.

| Cases | Methods | Energy | *f* | CI | Dmax | Dmax_oar | Dmean_oar |
|---|---|---|---|---|---|---|---|
| **Abdomen** | CONV | 50 | 2.23 | 0.90 | 110.8% | 108.1% | 49.1% |
| | ELO | 30 | 2.70 | 0.86 | 115.4% | 112.5% | 51.1% |
| | | 25 | 3.12 | 0.84 | 120.6% | 114.3% | 50.3% |
| | | 20 | 3.83 | 0.83 | 130.7% | 114.7% | 50.7% |
| **Brain** | CONV | 78 | 2.21 | 0.72 | 113.1% | 97.7% | 25.5% |
| | ELO | 40 | 2.79 | 0.69 | 117.8% | 100.5% | 26.3% |
| | | 30 | 3.82 | 0.66 | 120.1% | 100.0% | 29.7% |
| | | 20 | 6.34 | 0.61 | 125.9% | 104.2% | 29.7% |
| **Lung** | CONV | 100 | 1.15 | 0.74 | 113.1% | 106.8% | 16.9% |
| | ELO | 80 | 1.38 | 0.72 | 115.1% | 106.8% | 17.8% |
| | | 70 | 1.54 | 0.70 | 118.4% | 108.1% | 18.2% |
| | | 60 | 1.86 | 0.69 | 122.5% | 109.1% | 19.0% |

Table 3. **CONV *v.s.* ELO plans recalculated in RayStation.** The plan-quality dosimetric quantities from left to right: number of used energies, optimization objective value *f*, conformity index (CI), max target dose Dmax, mean OAR dose Dmean_oar and max OAR dose Dmax_oar.

| Cases | Methods | Energy | *f* | CI | Dmax | Dmax_oar | Dmean_oar |
|---|---|---|---|---|---|---|---|
| Abdomen | CONV | 50 | 2.42 | 0.84 | 112.9% | 109.6% | 48.1% |
| | ELO | 30 | 3.06 | 0.82 | 119.3% | 113.4% | 50.3% |
| | | 25 | 3.82 | 0.81 | 124.8% | 104.0% | 49.7% |
| | | 20 | 5.26 | 0.79 | 136.1% | 118.9% | 50.3% |
| Brain | CONV | 78 | 2.88 | 0.64 | 116.1% | 102.7% | 26.0% |
| | ELO | 40 | 3.44 | 0.62 | 119.8% | 100.0% | 24.8% |
| | | 30 | 4.89 | 0.60 | 124.6% | 99.4% | 28.2% |
| | | 20 | 6.54 | 0.58 | 124.6% | 101.5% | 28.3% |
| Lung | CONV | 100 | 1.42 | 0.72 | 113.9% | 104.0% | 15.4% |
| | ELO | 80 | 1.61 | 0.69 | 116.1% | 105.5% | 15.9% |
| | | 70 | 1.89 | 0.67 | 118.7% | 107.6% | 16.3% |
| | | 60 | 2.25 | 0.64 | 122.6% | 110.0% | 17.6% |

Table 4. **Delivery time.** The actual delivery time on the clinical machine (the unit is seconds).

| Cases | Energy layers | Delivery time |
|---|---|---|
| **Abdomen** | 50 | 110 |
| | 30 | 90 |
| **Brain** | 78 | 105 |
| | 40 | 54 |
| **Lung** | 100 | 178 |
| | 70 | 127 |

## 4. Discussion

This paper introduces and experimentally validates an IH-TPS with greedy energy layer optimization for fast IMPT. Experiments on abdomen, brain and lung cases demonstrate the accuracy and efficiency of the proposed IH-TPS. Compared to the standard IMPT plan, the treatment delivery time for brain and lung cases using ELO-based IMPT via IH-TPS was reduced by 48.6% and 28.7%, respectively.

As shown in Table 1, when plans generated by IH-TPS are recalculated in RayStation, the plan quality for the abdomen and brain cases were slightly degraded. It may be attributed to the difference in dose engine, i.e., the pencil beam algorithm in IH-TPS and the MC method in RayStation. This also leads to a reduced capability in decreasing the number of energy layers. For example, in the abdomen case, the number of energy layers can be reduced from 50 to 25 in IH-TPS, but only to 30 after recalculating in RayStation.

As shown in Table 4, the number of energy layers was reduced by 40% in the abdomen case, but the delivery time decreased by less than 20%. This limited reduction in the total delivery time by only reducing number of energy layers shows that the other time components (i.e., dose delivery time and spot scanning time) need to be accounted for as well. In the future work, we will integrate a comprehensive modeling of total delivery time [8,9] into IH-TPS to accurately model and optimize the total delivery time.

The proposed greedy method for ELO may be applicable to other inverse optimization problems in RT, such as the MMU problem [33,34] and beam angle optimization [35,36]. For example, when the high-dose-rate delivery is desirable (e.g., for the purpose of fast IMPT or FLASH [37,38]), it was shown in [33] that a modified orthogonal matching pursuit method can mitigate the failure of conventional optimization algorithms such as ADMM in handling the large-MMU problem and improve the plan quality.

## 5. Conclusion

An IH-TPS is developed and integrated with a novel ELO algorithm, which has been developed and experimentally tested to enable rapid IMPT with improved delivery efficiency while maintaining the quality of the treatment plan.

**References**

[1] Lomax, A. J., Bortfeld, T., Goitein, G., Debus, J., Dykstra, C., Tercier, P. A., Mirimanoff, R. O., 1999. A treatment planning inter-comparison of proton and intensity modulated photon radiotherapy. Radiotherapy and Oncology, 51(3), 257-271.

[2] Lomax AJ, 2008. Intensity modulated proton therapy and its sensitivity to treatment uncertainties 1: the potential effects of calculational uncertainties. Physics in Medicine & Biology, 53(4), 1027.

[3] Lomax AJ, 2008. Intensity modulated proton therapy and its sensitivity to treatment uncertainties 2: the potential effects of inter-fraction and inter-field motions. Physics in Medicine & Biology, 53(4), 1043.

[4] Li, H., Zhu, X. R., Zhang, X., 2015. Reducing dose uncertainty for spot-scanning proton beam therapy of moving tumors by optimizing the spot delivery sequence. International Journal of Radiation Oncology* Biology* Physics, 93(3), 547-556.

[5] Suzuki, K., Palmer, M. B., Sahoo, N., Zhang, X., Poenisch, F., Mackin, D. S., Lee, A. K., 2016. Quantitative analysis of treatment process time and throughput capacity for spot scanning proton therapy. Medical Physics, 43(7), 3975-3986.

[6] Pedroni E, Bacher R, Blattmann H, Böhringer T, Coray A, Lomax A, Lin S, Munkel G, Scheib S, Schneider U, Tourovsky A, 1995. The 200 - MeV proton therapy project at the Paul Scherrer Institute: Conceptual design and practical realization. Medical Physics, 22(1), 37-53.

[7] Paganetti H., 2018. Proton Therapy Physics. CRC press.

[8] Zhao L, Liu G, Zheng W, Shen J, Lee A, Yan D, Deraniyagala R, Stevens C, Li X, Tang S, Ding X, 2022. Building a precise machine-specific time structure of the spot and energy delivery model for a cyclotron-based proton therapy system. Physics in Medicine & Biology, 67(1), 01NT01.

[9] Zhao L, Liu G, Chen S, Shen J, Zheng W, Qin A, Yan D, Li X, Ding X, 2022. Developing an accurate model of spot-scanning treatment delivery time and sequence for a compact superconducting synchrocyclotron proton therapy system. Radiation Oncology, 17(1), 87.

[10] Van De Water, S., Kooy, H. M., Heijmen, B. J., Hoogeman, M. S., 2015. Shortening delivery times of intensity modulated proton therapy by reducing proton energy layers during treatment plan optimization. International Journal of Radiation Oncology* Biology* Physics, 92(2), 460-468.

[11] Cao, W., Lim, G., Liao, L., Li, Y., Jiang, S., Li, X., Zhang, X., 2014. Proton energy optimization and reduction for intensity-modulated proton therapy. Physics in Medicine & Biology, 59(21), 6341.

[12] Jensen, M. F., Hoffmann, L., Petersen, J. B. B., Møller, D. S., Alber, M., 2018. Energy layer optimization strategies for intensity-modulated proton therapy of lung cancer patients. Medical Physics, 45(10), 4355-4363.

[13] Lin, Y., Clasie, B., Liu, T., McDonald, M., Langen, K. M., Gao, H., 2019. Minimum-MU and sparse-energy-layer (MMSEL) constrained inverse optimization method for efficiently deliverable PBS plans. Physics in Medicine & Biology, 64(20), 205001.

[14] Lin, B., Li, Y., Liu, B., Fu, S., Lin, Y., Gao, H., 2024. Cardinality-constrained plan-quality and delivery-time optimization method for proton therapy. Medical Physics, 51(7), 4567-4580.

[15] Hong, L., Goitein, M., Bucciolini, M., Comiskey, R., Gottschalk, B., Rosenthal, S., Serago, C. and Urie, M., 1996. A pencil beam algorithm for proton dose calculations. Physics in Medicine & Biology, 41(8), p.1305.

[16] Bortfeld, T., 1997. Clinically relevant intensity modulation optimization using physical criteria. In XII International Conference on the Use of Computers in Radiation Therapy. Medical Physics Publishing.

[17] Wu, Q., Mohan, R., 2000. Algorithms and functionality of an intensity modulated radiotherapy optimization system. Medical Physics, 27(4), 701-711.

[18] Esser, E., Zhang, X. and Chan, T.F., 2010. A general framework for a class of first order primal-dual algorithms for convex optimization in imaging science. SIAM Journal on Imaging Sciences, 3(4), pp.1015-1046.

[19] Gabay D, Mercier B., 1976. A dual algorithm for the solution of nonlinear variational problems via finite element approximation. Computers & Mathematics with Applications, 2(1), 17-40.

[20] Glowinski R, Marroco A., 1975. On the approximation by finite elements of order one, and resolution, penalisation-duality for a class of nonlinear Dirichlet problems[J]. ESAIM: Mathematical Modelling and Numerical Analysis-Mathematical Modelling and Numerical Analysis, 9(R2), 41-76.

[21] Boyd, S., Parikh, N., Chu, E., Peleato, B., Eckstein, J., 2011. Distributed optimization and statistical learning via the alternating direction method of multipliers. Foundations and Trends® in Machine Learning, 3(1), 1-122.

[22] Goldstein T, Osher S., 2009. The split Bregman method for L1-regularized problems. SIAM Journal on Imaging Sciences, 2(2), 323-343.

[23] Gao, H., 2016. Robust fluence map optimization via alternating direction method of multipliers with empirical parameter optimization. Physics in Medicine & Biology, 61(7), 2838.

[24] Wang, Y., Yin, W. and Zeng, J., 2019. Global convergence of ADMM in nonconvex nonsmooth optimization. Journal of Scientific Computing, 78, 29-63.

[25] Eldar, Y. C., and Kutyniok, G, 2012. A mathematical introduction to compressive sensing. Cambridge University Press.

[26] Eldar, Y. C., Kuppinger, P., Bolcskei, H., 2010. Block-sparse signals: Uncertainty relations and efficient recovery. IEEE Transactions on Signal Processing, 58(6), 3042-3054.

[27] Zhu, Y. N., Zhang, X., Lin, Y., Lominska, C., Gao, H., 2023. An orthogonal matching pursuit optimization method for solving minimum-monitor-unit problems: applications to proton IMPT, ARC and FLASH. Medical physics, 50(8), 4710-4720.

[28] Gao H, 2019. Hybrid proton-photon inverse optimization with uniformity-regularized proton and photon target dose. Physics in Medicine & Biology, 64(10), 105003.

[29] Gao H, Lin B, Lin Y, Fu S, Langen K, Liu T, Bradley J, 2020. Simultaneous dose and dose rate optimization (SDDRO) for FLASH proton therapy. Medical Physics, 47(12), 6388-95.

[30] H. Gao, J. Liu, Y. Lin, G. Gan, G. Pratx, F. Wang, K. Langen, J. Bradley, R. Rotondo, H. Li, and R. C. Chen, 2022. Simultaneous dose and dose rate optimization (SDDRO) of the FLASH effect for pencil-beam-scanning proton therapy. Medical Physics, 49, 2014-2025.

[31] W. Li, Y. Lin, H. Li, R. Rotondo, and H. Gao, 2023. An iterative convex relaxation method for proton LET optimization. Physics in Medicine & Biology, 68, 055002.

[32] Zhang W, Li W, Lin Y, Wang F, Chen RC, Gao H, 2023. TVL1-IMPT: optimization of peak-to-valley dose ratio via joint total-variation and L1 dose regularization for spatially fractionated pencil-beam-scanning proton therapy. International Journal of Radiation Oncology* Biology* Physics, 115(3), 768-78.

[33] Low DA, Harms WB, Mutic S, Purdy JA, 1998. A technique for the quantitative evaluation of dose distributions. Medical Physics, 25(5), 656-61.

[33] Zhu, Y. N., Zhang, X., Lin, Y., Lominska, C., Gao, H., 2023. An orthogonal matching pursuit optimization method for solving minimum-monitor-unit problems: applications to proton IMPT, ARC and FLASH. Medical physics, 50(8), 4710-4720.

[34] Cai, J. F., Chen, R. C., Fan, J., Gao, H., 2022. Minimum-monitor-unit optimization via a stochastic coordinate descent method. Physics in Medicine & Biology, 67(1), 015009.

[35] Kaderka, R., Liu, K. C., Liu, L., VanderStraeten, R., Liu, T. L., Lee, K. M., Chang, C., 2022. Toward automatic beam angle selection for pencil-beam scanning proton liver treatments: A deep learning–based approach. Medical Physics, 49(7), 4293-4304.

[36] Shen, H., Zhang, G., Lin, Y., Rotondo, R. L., Long, Y., Gao, H., 2023. Beam angle optimization for proton therapy via group-sparsity based angle generation method. Medical Physics, 50(6), 3258-3273.

[37] Y. Lin, B. Lin, S. Fu, M. Folkerts, E. Abel, J. Bradley, and H. Gao, 2021. SDDRO-Joint: simultaneous dose and dose rate optimization with the joint use of transmission beams and Bragg peaks for FLASH proton therapy. Physics in Medicine & Biology, 66, 125011.

[38] J. Ma, Y. Lin, M. Tang, Y. Zhu, G. N. Gan, R. L. Rotondo, R. C. Chen, and H. Gao, 2024. Simultaneous dose and dose rate optimization via dose modifying factor modeling for FLASH effective dose. Medical Physics, 51, 5190-5203.